\documentclass[print,twocolumn]{revtex4}
\usepackage{amsmath}
\usepackage{pifont}
\usepackage{amssymb}
\usepackage{tipa}
\usepackage{amsfonts}
\usepackage{mathrsfs}
\usepackage{graphicx}
\usepackage{dcolumn}
\usepackage{bm}
\usepackage{txfonts}
\UseRawInputEncoding

\begin{document}
\title{Electric-Field Modulated Optical Transitions in Monolayer CrI$_3$ and Its Nanoribbons \footnote{Correspondence to: G H Zhou, ghzhou@hunnu.edu.cn}}
\author{Xianzhe Zhu$^1$}
\author{Pu Liu$^2$}
\author{Wence Ding$^2$}
\author{Benhu Zhou$^1$}
\author{Xiaoying Zhou$^3$}
\author{Guanghui Zhou$^{1,3}$}
\affiliation{$^1$Department of Physics, Shaoyang University, Shaoyang 422000, China}
\affiliation{$^2$School of Microelectronics and Physics, Hunan University of Technology and Business, Changsha 410205, China}
\affiliation{$^3$Hunan Provincial Key Laboratory of Intelligent Sensors and Advanced Sensor Materials,
School of Physics and Electronics Science, Hunan University of Science and Technology, Xiangtan 411201, China}
\affiliation{$^4$Department of Physics and Key Laboratory for
Low-Dimensional Structures and Quantum Manipulation (Ministry of
Education), Hunan Normal University, Changsha 410081, China}


\begin{abstract}
\textbf{Abstract}: The successful synthesis of few-layer CrI$_3$ has opened new avenues for research in two-dimensional magnetic
materials. Owing to its simple crystal structure and excellent physical properties, layered CrI$_3$ has been extensively studied in
magneto-optical effects, excitons, tunneling transport, and novel memory devices. However, the most current theoretical studies
rely heavily on the first-principles calculations, and a general analytical theoretical framework, particularly for electric-field
modulation and transport properties, is still lacking. In this work, using a 28-band tight-binding model combined with linear
response theory, we systematically investigate the optoelectronic response for monolayer CrI$_3$ and its nanoribbons. The results
demonstrate that: (1) a vertical electric field can selectively close the band gap of one spin channel while the other remains
insulating, resulting a transition to an half-metallic state; (2) the electric field dynamically shifts the optical transition
peaks, providing a theoretical basis for extracting band parameters from experimental photoconductivity spectra; (3) nanoribbons
with different edge morphologies exhibit distinct edge-state distributions and electronic properties, indicating that optical
transition can be dynamically modualted through edge design. The theoretical model developed in this study, which can describe
external electric field effect, offers an efficient and flexible approach for analytically investigating the CrI$_3$ family and
related materials. This model overcomes the limitations of first-principles methods and provides a solid foundation for designing
spintronic and optoelectronic devices controlled by electric fields and edge effect.

\end{abstract}
\maketitle

\section{Introduction}
As is well known, the Mermin-Wagner theorem states that long-range magnetic order is difficult to maintain in two-dimensional (2D) systems at finite temperatures due to thermal fluctuations  \cite{Mermin1966}, therefore research on 2D magnetic materials remained largely unexplored for an extended period. However, sufficiently strong magnetic anisotropy can overcome the effects of thermal disturbances. The introduction of magnetocrystalline anisotropy, magnetoelastic anisotropy, or magnetostress anisotropy makes the realization of 2D magnetic materials possible \cite{Irkhin1999}.

In 2016, several research groups successfully synthesized monolayer and few-layer NiPS$_3$ \cite{Cheng2016}, FePS$_3$ \cite{Lee2016,Wang2016} and CrSiTe$_3$ \cite{Lin2015}, and demonstrated the existence of long-range magnetic order through Raman spectroscopy. This discovery marked substantial progress in the field of 2D magnetic materials. In 2017, researchers confirmed the presence of intrinsic magnetism in few-layer Cr$_2$Ge$_2$Te$_6$ \cite{Gong2017} and monolayer CrI$_3$ \cite{Huang2017}. These findings not only opened new directions for research on 2D magnetic materials but also provided novel insights for the development of innovative spintronic devices \cite{Huang2017,Jin2018,Kong2019,Yu2019,Frey2018,Kargar2020,Deng2020}.

Bulk CrI$_3$ is a layered transition metal halide with a Curie-temperature of approximately 61 K, exhibiting strong magnetic anisotropy and an out-of-plane easy magnetization axis \cite{Dillon1965,McGuire2015}. Consequently, monolayer and few-layer CrI$_3$ can be obtained experimentally through mechanical exfoliation, with their magnetic properties showing significant dependence on layer thickness. Furthermore, polar magneto-optical Kerr effect measurements have confirmed that monolayer CrI$_3$ is an intrinsic Ising ferromagnet with a Curie-temperature of about 45 K \cite{Huang2017,Lado2017,Wang2018,Huang2020}.

Due to its simple crystal structure and excellent physical properties, CrI$_3$ layer has been extensively studied in various fields including magneto-optical effects \cite{Huang2020,Seyler2017,Sun2019}, excitons \cite{Jin2020,Mukherjee2020}, and tunneling transport \cite{Klein2018,Li2019,Wang2019}. The magnetic states of few-layer CrI$_3$ are highly sensitive to external perturbations, and various approaches such as applied electric field \cite{Wang2018b,Burch2018,Webster2018,Jiang2018}, electrostatic doping \cite{Huang2018,Jiang2018b}, and strain engineering \cite{Webster2018b,Baidya2018} can effectively modulate its electronic structure and magnetic properties, providing new methods to enhance the magnetic stability. For instance, constructing heterostructures of CrI$_3$ with other materials can modify the band gap and magnetization, enabling the realization of valley-polarized quantum anomalous Hall effect and manipulation on band valley and electron spin \cite{Zhang2018,Zhang2019}. External magnetic field can also precisely control the direction of magnetic moments for the system, inducing magneto-photoluminescence \cite{Jiang2018c}. Further, an electric field can switch monolayer CrI$_3$ between ferromagnetic and antiferromagnetic states \cite{Jiang2018}. And applied strain regulates the magnetic anisotropy and topological properties of CrI$_3$, leading to changes in magneto-optical Kerr signals \cite{Webster2018b,Baidya2018}. Additionally, defects and adsorbed atoms can significantly alter the electronic structure and magnetic properties of monolayer CrI$_3$ \cite{Zhao2018,Guo2018}, resulting in modulation of the band gap size, transformation from indirect to direct band gap semiconductors, and modification of magnetic moment magnitudes.

Furthermore, research on the optical properties of 2D magnetic materials is of great significance for their applications in optoelectronic devices. Experimental techniques such as photoluminescence spectroscopy and Raman scattering spectroscopy are commonly used to characterize the optical properties of materials. Raman spectroscopy studies on the lattice vibrations revealed that few-layer CrI$_3$ single crystals exhibits monoclinic stacking in the high-temperature phase and rhombohedral stacking in the low-temperature phase \cite{Song2019,Niu2020}. Manipulating the magnetic order of few-layer CrI$_3$ crystals can control Raman optical selection rules, providing opportunities to explore new magneto-optical effects and spin-phonon coupling mechanisms \cite{Zhang2019b}. Through photoluminescence spectroscopy, scientists have discovered that spontaneous magnetization exists within few-layer CrI$_3$ crystals at low temperatures, resulting in significant differences in fluorescence intensity generated by transitions of two spin electrons \cite{Seyler2017}.

On the theoretical work side, it has been found that all-optical magnetization switching can be achieved in single-layer CrI$_3$ using circularly polarized light \cite{Zhang2022}. The exciton states, light absorption, and magneto-optical properties of single-layer CrI$_3$ can also be regulated by strain \cite{Wu2019,Wang2025}. In addition. layered CrI$_3$ exhibits strong Kerr and Faraday rotations in the visible to ultraviolet range, comparable to classical magneto-optical materials \cite{Tang2022}. Additionally, the electronic states near the Fermi level are spin-polarized in CrI$_3$. These excellent optical and magneto-optical properties indicate that layered CrI$_3$ has broad application prospects in nano-optoelectronic and spintronic devices.

Current research on the optical properties of CrI$_3$ mainly focuses on experimental characterization (such as magneto-optical spectroscopy, Raman scattering) and first-principles calculations (such as band structure and magnetic ground state analysis, etc.). These works have successfully revealed its magnetic-order-dependent optical response characteristics. However, calcuating the optical conductivity from first principles involves handling numerous \emph{k}-points and interband transitions, limiting rapid exploration of dimensional effect. This lack of efficiency, combined with the absence of a transparent physical picture, obstructs an intuitive understanding of the underlying transport mechanisms. The development of an analytical theoretical model for monolayer CrI$_3$ faces two primary challenges. First, its indirect bandgap, with band edges located at distinct \emph{k}-points, and its completely spin-polarized bands make it difficult for a single kp model to simultaneously describe both spin channels. Second, the ferromagnetic order breaks time-reversal symmetry, which significantly reduces the effective symmetry of the system from that of its crystallographic space group. Consequently, constructing a k¡¤p model requires numerous parameters that are difficult to determine. Although monolayer CrI$_3$ possesses an inherent structural symmetry described by its space group, the emergence of ferromagnetic order breaks the time-reversal symmetry, thereby significantly reducing the system's actual effective symmetry. Hence the \emph{k¡¤p} model has many parameters that are difficult to be determined.

Some theoretical works have constructed a tight-binding (TB) model based on the 3\emph{d}-orbitals of magnetic atom Cr and the 5\emph{p}-orbitals of I atoms, with which the band structure of monolayer CrI$_3$ is successfully fitted \cite{Yananose2022,Rozhansky2024}. This result can approximately describe the band structure throughout the first Brillouin zone and has been used to analyze the magnetic and topological states of bilayer and twisted CrI$_3$. Therefore, in our work, based on the TB mode and linear response theory, we further investigate the spin-polarized band structures, edge states and band modulation under the vertical electirc field \cite{Yananose2022}. At the same time, we predict the optoelectronic property of one-dimensional CrI$_3$ nanoribbons and explores the influence of different edge morphologies on optical transition behavior. Our results may provide a theoretical basis for designing nanoscale optical modulators and spin filters based on the few-layer CrI$_3$ materials.

\section{Model and Method}
\subsection{Tight-binding model for monolayer CrI$_3$}

\begin{figure*}
\includegraphics[width=2\columnwidth]{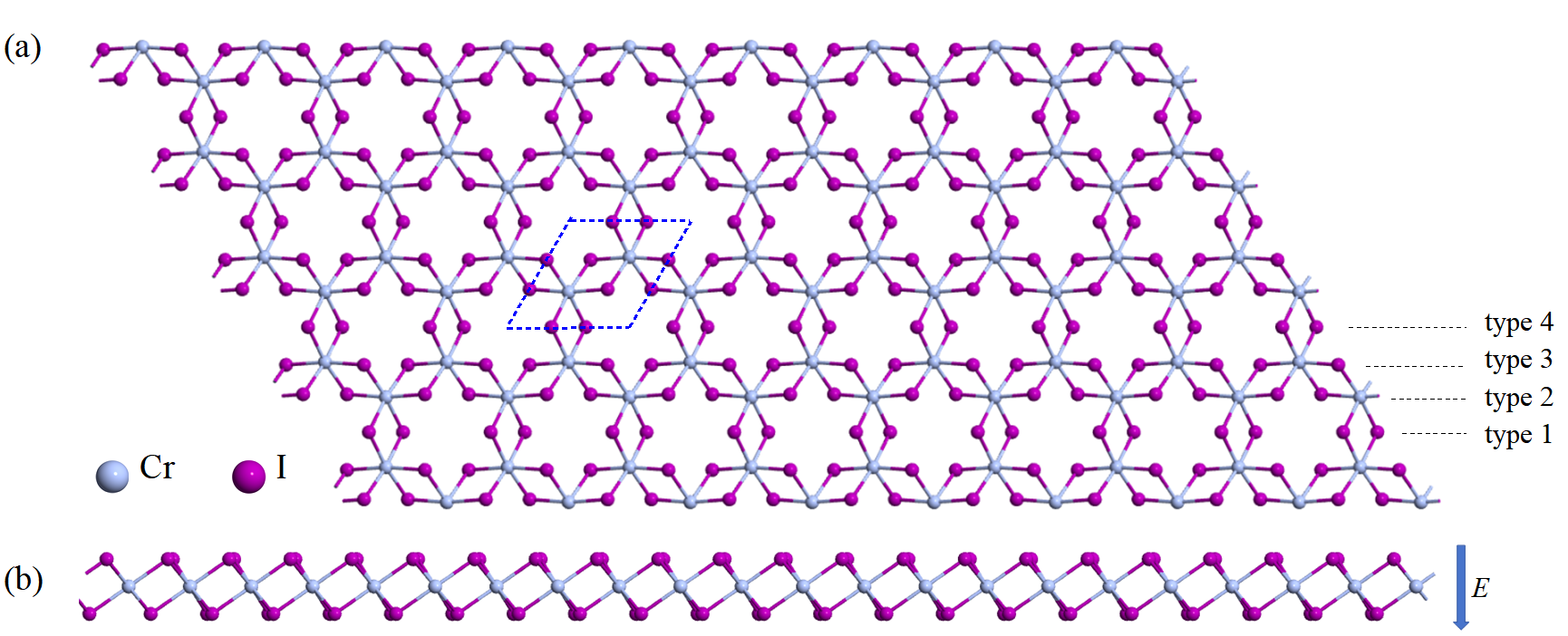}
\caption
{Schematic illustrations of (a) the primitive cell of monolayer CrI$_3$, (b) the Brillouin zone, and (c) the atomic structures of CrI$_3$ nanoribbons with the four distinct boundary types marked in which are defined by different edge-terminating atoms. Different combinations of these boundary types form nanoribbons with various edge configurations, as detailed in the main text. Cr and I atoms are represented by light purple and deep purple spheres, respectively.}
\label{Fig.1}
\end{figure*}

Monolayer CrI$_3$ has a crystal structure belonging to the space group P$\overline{3}$1m. Its structure comprises a I-Cr-I trilayer where Cr atoms form a hexagonal honeycomb lattice. Each Cr atom is octahedrally coordinated by six nearest-neighbor I atoms, with Cr at the octahedral center and I at the vertices, as depicted in Fig. 1 (light purple Cr and deep purple I). The in-plane lattice constant is $a$ = 6.98525 ${\AA}$. Figure 1(a) shows the unit cell, defined by primitive vectors $\mathbf{a}=a(1,0,0)$, $\mathbf{b}=a(1/2, \sqrt{3}/2,0)$, and $\mathbf{c}=a(0,0,1)$, containing 2 Cr and 6 I atoms.

For an accurate description of the electronic structure and physical properties of monolayer CrI$_3$, a 28-band TB model has been developed in previous work \cite{Yananose2022},
employing the five \textit{d}-orbitals of the Cr atom $ \{|d_{z^2}\rangle, |d_{x^2-y^2}\rangle, |d_{xy}\rangle, |d_{xz}\rangle, |d_{yz}\rangle \}$ and the three \textit{p}-orbitals of the I atom $\{|p_x\rangle, |p_y\rangle, |p_z\rangle \}$ as the basis set. In our calculations, we neglect spin-orbit coupling and express the Hamiltonian in the following matrix form
\begin{equation}
\begin{aligned}
H = \begin{pmatrix} H_{\uparrow} & 0 \\ 0 & H_{\downarrow} \end{pmatrix}.
\end{aligned}
\end{equation}
Here, $H_{\uparrow}$ and $H_{\downarrow}$ represent the hopping matrices for spin-up and spin-down electrons, respectively. The matrix elements between orbital $\alpha$ on atom $i$ and orbital $\beta$ on atom $j$ at lattice vector \textbf{k} are
\begin{equation}
(H_k)_{\sigma i\alpha,\sigma'j\beta}=\sum_{\mathbf{R}} e^{i\mathbf{k}\cdot(\mathbf{R}+\mathbf{r}_j-\mathbf{r}_i)} \left\langle \varphi_{i\alpha}^{\sigma}(\mathbf{r}-\mathbf{r}_i) \middle| \tilde{H}\middle|\varphi_{j\beta}^{\sigma'}(\mathbf{r}-\mathbf{R}-\mathbf{r}_j) \right\rangle,
\end{equation}
where $\langle \varphi_{i\alpha}^{\sigma}(\mathbf{r}-\mathbf{r}_i)|\hat{H}|\varphi_{j\beta}^{\sigma^{\prime}}(\mathbf{r}-\mathbf{R}-\mathbf{r}_j)\rangle$
represents the hopping energy of an electron with spin $\sigma$ in the $\alpha$-orbital of atom $i$ at position $\mathbf{r}_i$ to one
with spin $\sigma^{\prime}$ in the $\beta$-orbital of atom $j$ at position $\mathbf{r}_j$, in which $\mathbf{R}$ is the lattice vector.
The subscripts $(\sigma i\alpha)$ denote the associate spin, atom, and orbital, respectively. The hopping parameters are determined
using the Slater-Koster method \cite{Slater1954}, and other parameters in this work are taken from Ref. \cite{Yananose2022}.

This TB model is constructed to incorporate all key orbitals and their hybridizations via multiple hopping terms, enabling it to accurately capture the electronic band structure throughout the first Brillouin zone. The results show good in agreement with both experimental data and first-principles calculations, making the model suitable for further investigations of transport properties in monolayer CrI$_3$. Additionally, the model is well-suited for incorporating external perturbations (e.g., electric/magnetic fields, strain), paving the way for computational studies of tunable transport properties.

Based on this model, we further investigate the modulation on the electronic structure of monolayer CrI$_3$ under a perpendicular electric field. Owing to the trilayer structure of monolayer CrI$_3$, the effect of a perpendicular electric field can be incorporated into the TB model by adding a corresponding on-site energy term
\begin{equation}
H_{E_t} = E_t z_i I,
\label{eq:field_hamiltonian}
\end{equation}
where $E_t$ denotes the strength of the perpendicular electric field at the $z_i$-position of the \emph{i} atom, and $I$ is the identity matrix. Furthermore, we investigate the electronic structure and optical transitions of the monolayer CrI$_3$ nanoribbons with different boundary configurations. Monolayer CrI$_3$ nanoribbons with zigzag edges has four types edge terminations (type 1, 2, 3 and 4), which shown in the Fig. 1(c).  The four edge termination combinations correspond to four distinct configurations of zigzag CrI$_3$ naniribbon, which are Ribbon\_1 (boundary types 1 and 3), Ribbon\_2 (boundary types 2 and 3), Ribbon\_3 (both edges of boundary type 1), and Ribbon\_4 (both edges of boundary type 3).

By applying Bloch's theorem to the periodic nanoribbon structure, we obtain the difference Schr\"{o}dinger equation for the ribbon system \cite{Datta2005}
\begin{equation}
H = H_{00} + H_{01} e^{ik_x} + H_{01}^\dagger e^{-ik_x},
\label{eq:ribbon_hamiltonian}
\end{equation}
where $H_{00}$ ($H_{01}$) represents the intra-cell (inter-cell) hopping elements of the Hamiltonian, and $k_x$ denotes the wave vector along the ribbon direction.

To distinguish and investigate the influence of edge morphology on the electronic structure of the nanoribbons, we further calculate the local density of states (LDOS) for these four edge configurations. The LDOS reflects the spatial distribution characteristics of the wave functions, thereby effectively distinguishing the spatial localization behavior of edge states and the differences in electronic structures.

Within tour TB framework, the Green's function for atom $\alpha$ in the l-th row is given by
\begin{equation}
g_{l, \alpha}^r(\omega, k_x) = \frac{1}{L_x} \frac{\phi_{\alpha}(k_x, l) \phi_{\alpha}^*(k_x, l)}{\omega - E(k_x, l) + i \delta},
\label{eq:green_function}
\end{equation}
where $\delta$ is an infinitesimal positive number, and $\phi_{\alpha}(k_x, l)$ represents the wave function of atom $\alpha$ in the l-th row. For nanoribbons, the local density of states (LDOS) at the l-th row can be expressed as
\begin{equation}
\text{LDOS}(\omega) = -\frac{2}{\pi} \int_{-\pi/a}^{\pi/a} g_{l, \alpha}^r(\omega, k_x) dk_x,
\label{eq:ldos}
\end{equation}
where the integral is taken over the entire Brillouin zone in the $k_x$-direction.
The LDOS reflects the localization characteristics of the electronic states along the width direction
of the nanoribbon and can be used to analyze the contribution of electrons at arbitrary spatial
positions in the energy bands in reciprocal space. In this work, we utilize the LDOS to identify
the edge state characteristics of CrI$_3$ nanoribbons with different edge configurations.

\subsection{Optical conductivity}
To verify the aforementioned band structure and understand the experimentally observed optical properties of CrI$_3$, we also investigate its optical response. Under linearly polarized light illumination, electrons in the valence band can be excited and transition to the conduction band. This process typically occurs at the same wave vector positions and can be characterized by the optical absorption spectrum. Using linear response theory, the absorption spectrum can be calculated via the Kubo formula \cite{Ando1974, Koshino2008}

\begin{equation}
\begin{split}
\sigma_{xx}(\omega) = C \sum_{\nu,c,k} & [f(E_n) - f(E_m)] \times \\
& \left| \langle \phi_n(\mathbf{k}) | \hat{v}_x | \phi_m(\mathbf{k}) \rangle \right|^2 \delta [\hbar\omega - (E_n(\mathbf{k}) - E_m(\mathbf{k}))]
\end{split}
\label{eq:optical_conductivity}
\end{equation}
where $\sigma_{xx}$ represents the transverse optical conductivity in the 2D case, $C$ is a constant determined by the system parameters, and $n$($m$) denote the band index for the conduction (valence) bands, respectively. The velocity operator in the $x$-direction is given by $v_x = \frac{1}{\hbar} \frac{\partial H}{\partial k_x},\label{eq:velocity_operator}$
and $f(E_n)$ is the Fermi-Dirac distribution function. For one-dimensional nanoribbons, there is only one periodic direction (the $x$-direction), thus we only need to investigate their transverse optical conductivity.

\section{Results and Discussion}
\subsection{Optical transitions in monolayer CrI$_3$}
Using the tight-binding model from Eq. (1), we calculated the band structure of monolayer CrI$_3$ under electric field modulation, as shown in Fig.2. In the figure, the red and blue bands correspond to spin-up and spin-down electronic states, respectively. While the Hamiltonian has the same form for both spins, the associated hopping parameters are physically distinct for each channel \cite{Yananose2022}. The bands exhibit significant spin polarization characteristics, regardless of the presence of an external electric field.

\begin{figure}
\center
\includegraphics[width=1.0\columnwidth]{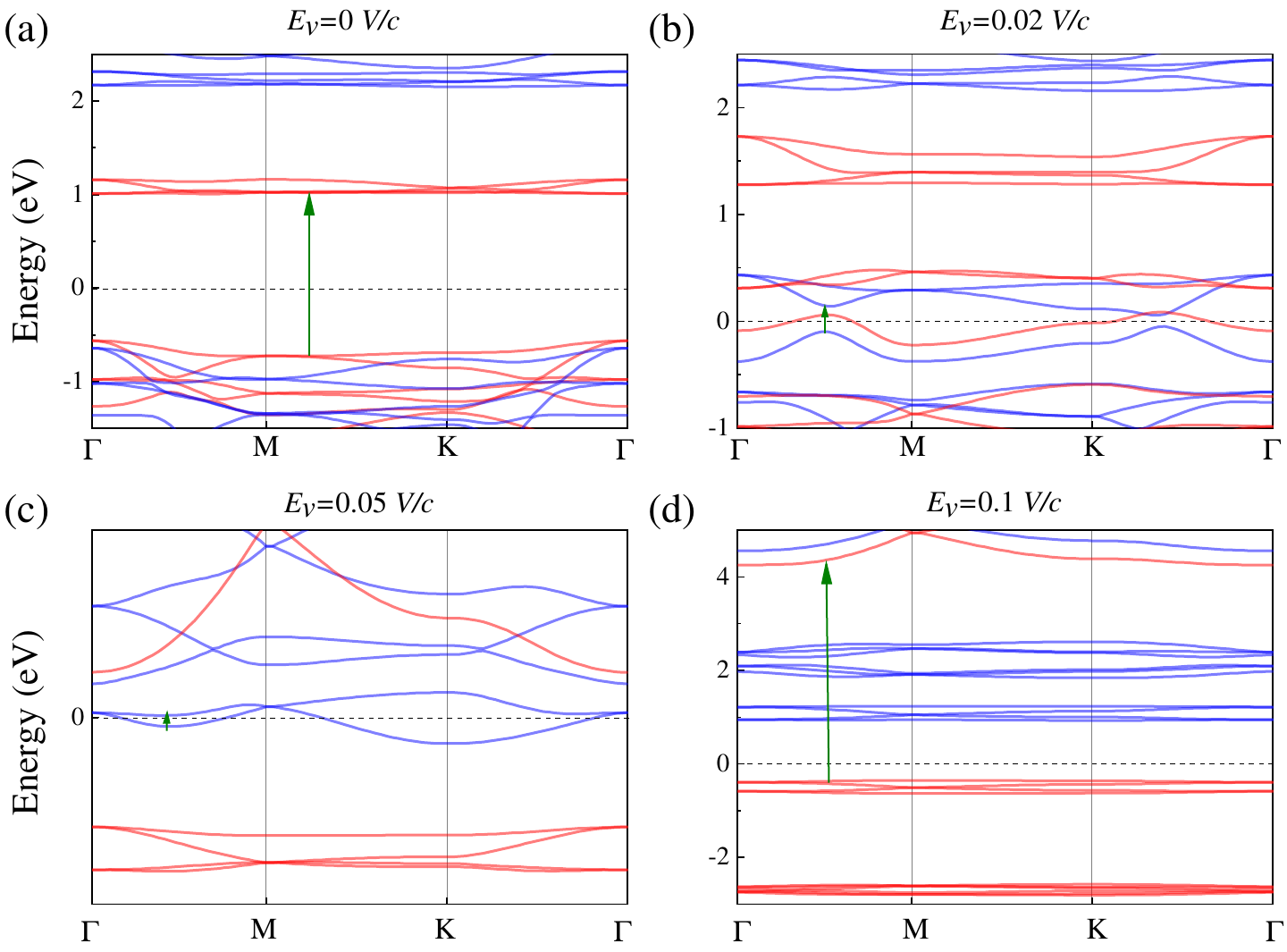}
\caption{Evolution of band structures in monolayer CrI$_3$ under applied vertical electric field. Red and blue curves denote spin-up and spin-down bands, respectively.}
\label{Fig.2}
\end{figure}

As shown in Fig. 2(a) without electric field, the bands near the Fermi level are primarily contributed by a single spin state, with a band gap of approximately 1.8 eV. The TB model results show good agreement with first-principles calculations across the entire first Brillouin zone. Under a weak vertical electric field, the two spin channels exhibit a divergent evolution: the spin-up gap monotonically decreases and closes at a critical field, inducing a half-metallic state, while the spin-down gap remains large and insulating. This distinct response provides a mechanism for electric-field-controlled spintronics.

Furthermore, under a weak electric field with $E_v$=0.02 V/$c$ (where $c$ represents the thickness of the 2D unit cell along the z-direction), the Fermi level at $E_F$=0 eV the system exhibits half-metallic behavior. In this state, it intersects exclusively the spin-up subbands, exhibiting half-metallic behavior as shown in Fig. 2(b). The energy range with 100\% spin polarization spans from -48 meV to 56 meV. When the electric field increases to $E_v$ = 0.05 V/$c$, the Fermi level crosses exclusively the spin-down (blue) subbands, with 100\% spin polarization maintained in the energy range of -140 meV to 130 meV. Additionally, a band gap of approximately 450 meV separates the states from the spin-up bands. Moreover, at $E_v=0.1$ V/$c$, the bands near the Fermi level are predominantly contributed by spin-down electronic states. Further, at a critical field of 0.1 V/$c$, the electronic states near the Fermi level become dominated by the spin-down channel. The energy window for 100\% spin polarization expands to 0.97$-$2.6 eV, nevertheless a band gap persists throughout this range.

\begin{figure}
\center
\includegraphics[width=1.0\columnwidth]{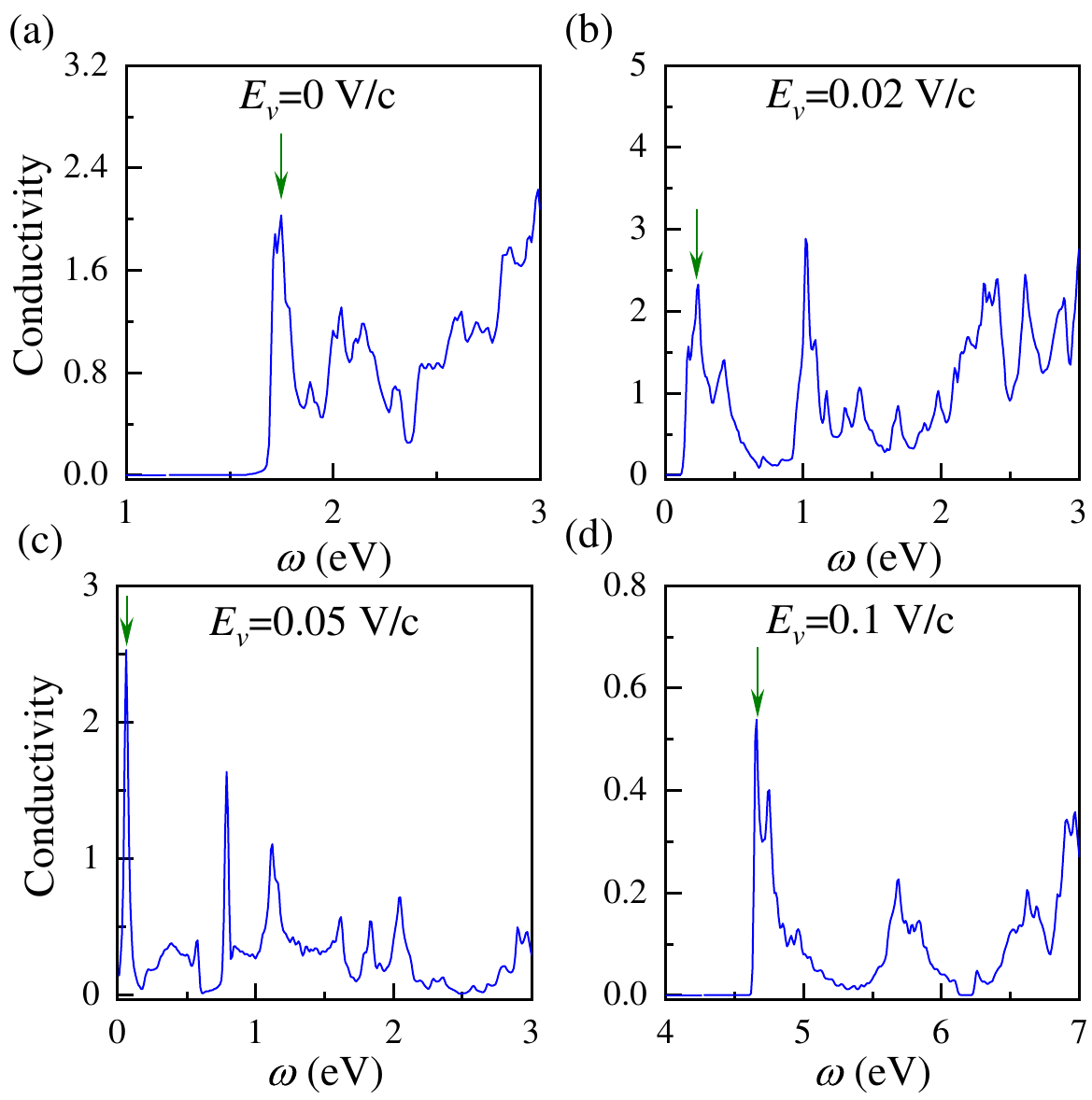}
\caption{(Color online) Optical conductivity of monolayer CrI$_3$ under different strengths of vertical electric field. The transition peaks marked by green arrows correspond to inter-subband transitions shown in Fig.2.}
\label{Fig.3}
\end{figure}

Since the electronic band structure is not directly accessible in experiments, its features are typically inferred from optical transition spectra. Therefore, based on the above-described band evolution, we have calculated the electric-field-tuned optical transition spectra for monolayer 2D CrI$_3$, which are presented in Fig. 3. Therefore, based on the aforementioned band evolution results, we further calculated the optical transition spectra of monolayer 2D CrI$_3$ under electric field modulation, as presented in Fig.3. The first transition peak (marked by green arrows) at different electric fields corresponds to the transition processes indicated by arrows in Fig.2. Since spin-orbit coupling effects are not considered in this work for CrI$_3$, optical transitions can only occur between subbands with identical spin configurations. In our calculations, the Fermi level is consistently set at $E_F= 0$ eV.

Under zero electric field conditions [Fig.3(a)], the first peak in the optical conductivity spectrum appears at $\hbar\omega = 1.74$ eV, corresponding to the transition marked by the green arrow in Fig.2(a), which originates from transitions between spin-up subbands. The lack of optical conductivity peaks below 1.74 eV implies a band gap of monolayer CrI3, which is apporximatlely 1.74 eV. This characteristic can be utilized in optical experiments to determine the band gap size of monolayer CrI$_3$. As the incident photon energy increases, more spin-up electrons from the valence band are excited and transition to the conduction band. However, the spin-down band gap remains too large for transitions to occur until the incident photon energy reaches $\hbar\omega > 2.7$ eV.

When a vertical electric field is applied, the band gap of the spin-up channel gradually closes. Although the band gap of the spin-down channel shows negligible variation, the overall energy position of these bands shifts upward. At $E_v = 0.02$ V/$c$, the bands near the Fermi level are primarily contributed by spin-down electronic states, resulting in the emergence of the first transition peak near zero energy, which originates from transitions between spin-down subbands. With the Fermi level set between these two edge states, the electronic structure within an energy range of about 3 eV around EF consists purely of spin-down character.

\begin{figure}
\center
\includegraphics[width=1.0\columnwidth]{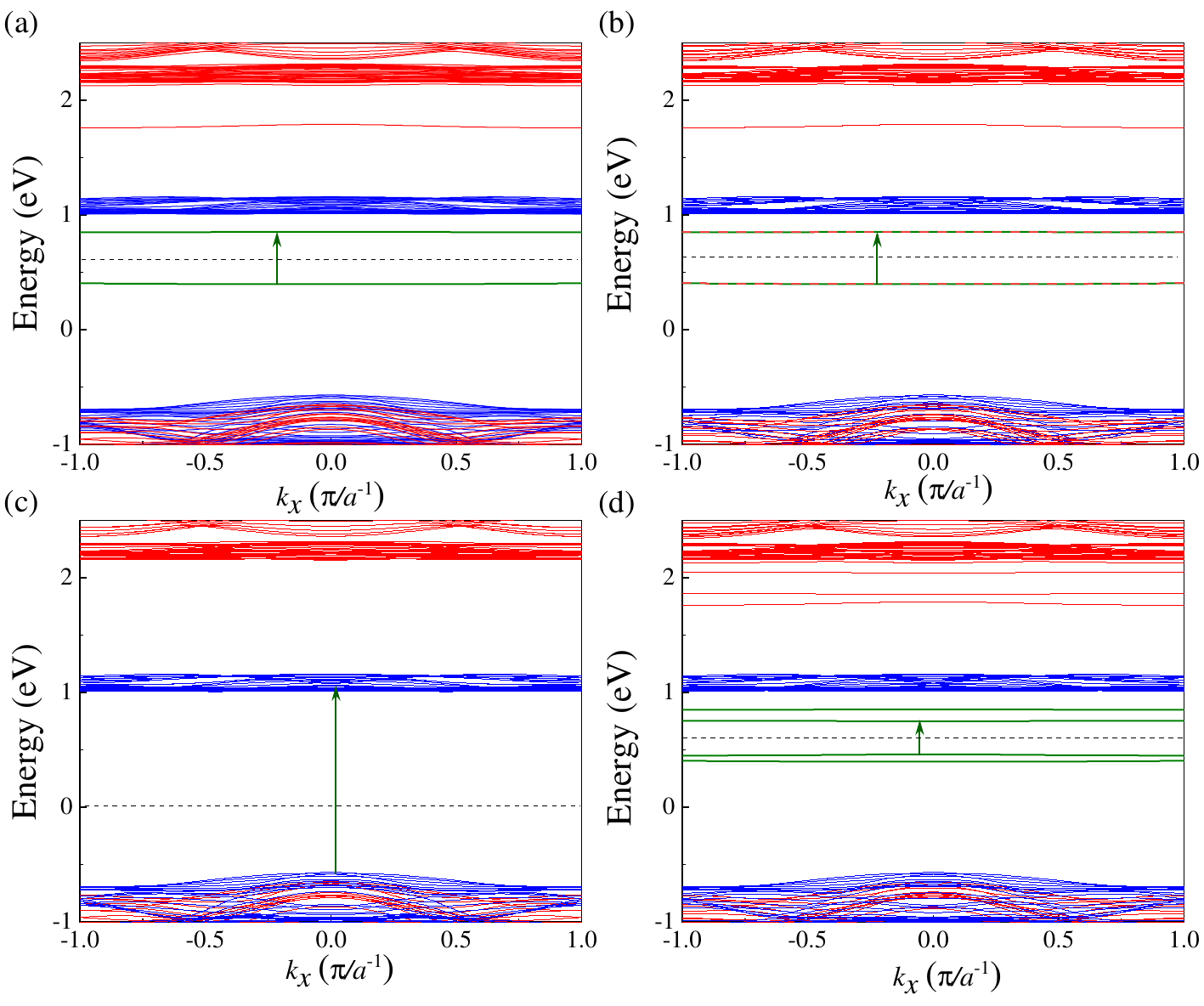}
\caption{(Color online) Band structures of monolayer CrI$_3$ nanoribbons with distinct edge configurations. Green solid lines and red dashed lines represent edge states, which are contributed by spin-down electronic states. The arrows in the figures correspond to the optical transition processes shown in Fig. 6.}
\end{figure}

At $E_v = 0.1$ V/$c$, the band gap in the spin-up channel reopens, and the electronic states near the Fermi level are no longer dominated by a single spin component. The electronic states below the Fermi level are dominated by spin-up contributions, while those above it are mainly spin-down. Since optical transitions between different spin states are forbidden in the absence of spin-orbit coupling, no transition peaks appear across a wide energy range ($\hbar\omega < 4.6$ eV). The first transition peak in Fig.3(d) originates from transitions between spin-up electronic states.

Therefore, by applying an external electric field, we can modulate the band structure to control both the spin composition near the Fermi level and the band gap, thereby enabling manipulation of optical transition energies and channels. This provides a theoretical foundation for designing electric-field-tunable spin-optoelectronic devices.

\subsection{Optical transitions for monolayer CrI$_3$ nanoribbons with different edge configurations}
Tailoring 2D CrI$_3$ into nanoribbons gives rise to distinct quantum confinement and edge effects, opening new avenues for property manipulation and potential applications in nanodevices. Therefore, based on the TB models in Eqs. (3) and (6), we also investigate the band structures and optical transition behaviors of CrI$_3$ nanoribbons with four different edge configurations.

Taking CrI$_3$ nanoribbons with a width of $N=5$ atomic rows as an example, we study the band characteristics of four edge configurations, as shown in Fig. 4. The band structures of Ribbon\_1 and Ribbon\_2 are highly similar, as shown in Figs. 4(a) and (b), which are composed of edge 1 and edge 3, and edge 2 and edge 3, respectively. Although their bulk bands are nearly identical, there are significant differences in the number of edge states: Ribbon\_1 has only two edge states, while Ribbon\_2 has four edge states, which are pairwise degenerate.

\begin{figure*}
\center
\includegraphics[width=1.6\columnwidth]{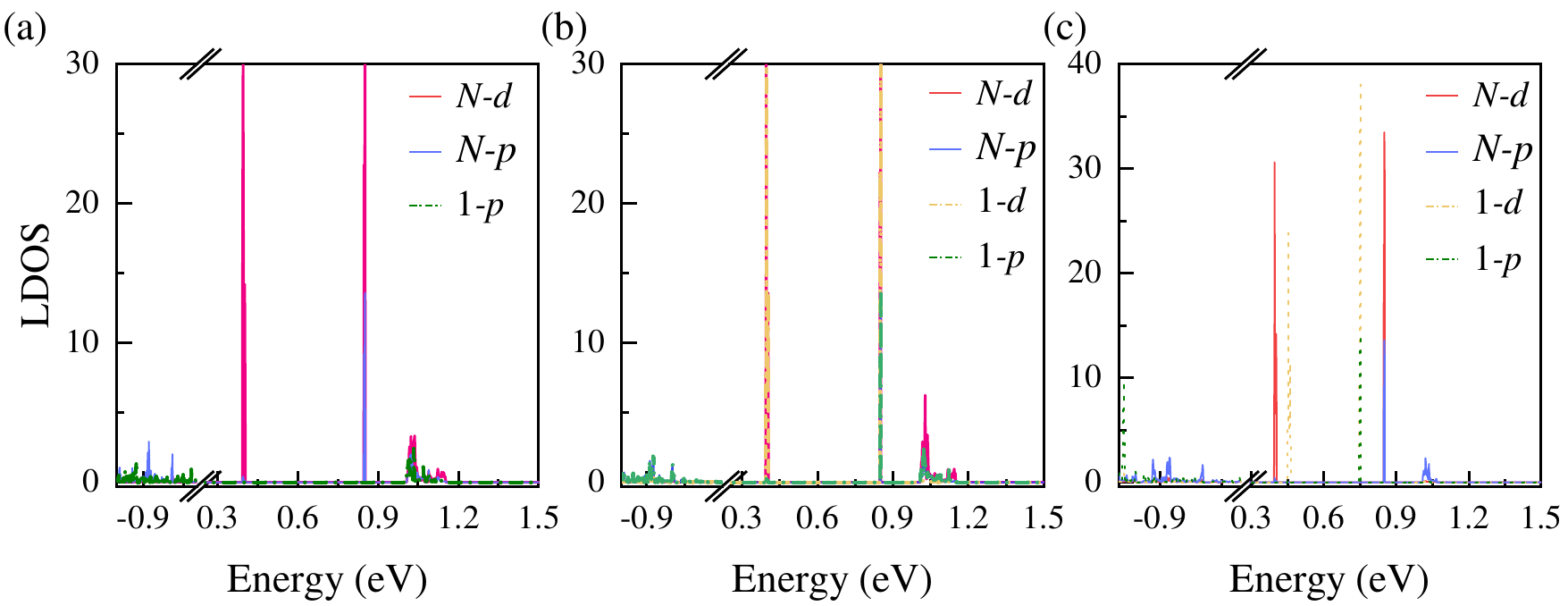}
\caption{The LDOS for edge atoms of monolayer CrI$_3$ nanoribbons: (a) Ribbon\_1, (b) Ribbon\_2, and (c) Ribbon\_4.}
\end{figure*}

To clearly display the degenerate edge states, we distinguish them with green solid lines and red dashed lines in Fig. 4. The edge states in both Ribbon\_1 and Ribbon\_2 are contributed by spin-down electronic states. With the Fermi level set between these two edge states, the electronic structure within an energy range of about 3 eV around $E_F$ consists purely of spin-down character. Structurally, the edge norphology on one side for Ribbon\_1 and 2 are the same, where Cr and I atoms alternate. Their other edges are different: Ribbon\_1 is terminated solely by I atoms (Edge 1), while Ribbon\_2 also has the alternating Cr-I structure (Edge 3). This structural difference suggests that the edge states primarily originate from Cr atoms rather than I atoms.

Figure 4(c) shows the band structure of Ribbon\_3, where both edges are of type 1, meaning that the atoms exposed on both sides are I atoms. The band structure near the Fermi level in this nanoribbon is still primarily contributed by spin-down electronic states. The absence of edge states in the band structure of Ribbon\_3 further supports the conclusion that these states do not originate from I atoms. Although Ribbon\_4 has Cr-terminated edges (Edge 3) on both sides, its structure remains asymmetric, as shown in Fig. 1(d). This configuration yields four non-degenerate edge states, all of spin-down character, which are plotted as green solid lines in Fig. 4(d). This type of edged nanoribbon also possesses four edge states, all contributed by spin-down electronic states, and these four edge states are non-degenerate, as indicated by the green solid lines in Fig. 4(d).

From the electronic band structures, it can be seen that edge states are present in three of the four monolayer CrI$_3$ nanoribbons with distinct edge configurations, with their number and degeneracy varying across the systems. To investigate the origins of these three distinct types of edge states, we calculated the local density of states (LDOS) for the edge atoms of Ribbon\_1, Ribbon\_2, and Ribbon\_4 based on Eq. (6), with the results presented in Fig. 5. Since the TB model used in this study fully considers the five \emph{d}-orbitals of Cr and the three \emph{p}-orbitals of I, it can accurately describe the electronic state distribution, thus enabling accurate analysis of the origins of various edge states through LDOS. Given that the electronic states near the Fermi level are primarily contributed by spin-down electrons, only the LDOS of spin-down electronic states is calculated in this work.

First, as shown in Fig. 5(a), there are two edge states in the electronic structure of Ribbon\_1, which mainly originate from the \emph{d}-orbitals of the Cr atoms in the \emph{N}-th row, with minor contributions from the \emph{p}-orbitals of the I atoms in the \emph{N}-th row. Ribbon\_2 [Fig. 5(b)] exhibits four edge states, with both edges consisting of alternating Cr and I atoms. For Ribbon\_2, as shown in the Fig. 5(b), has four edge states and features alternating Cr and I atoms on both edges. The edge states are primarily contributed by the \emph{d}-orbitals of Cr atoms at both edges, with equal contributions from the two edges, resulting in pairwise degenerate edge states. Ribbon\_4 also has edges composed of alternating Cr and I atoms, thus it also possesses four edge states. These four edge states mainly originate from the d orbitals of Cr atoms in the 1st and \emph{N}-th rows. When the Fermi level is set in the middle of the edge states, the two edge states near the Fermi level are mainly contributed by the \emph{d}-orbitals of Cr atoms in the 1st row, while the other two edge states are mainly contributed by the \emph{d}-orbitals of Cr atoms in the \emph{N}-th row. However, the structure of Ribbon\_4 is not completely symmetric, causing these four edge states to be split.
\begin{figure*}
\center
\includegraphics[width=1.6\columnwidth]{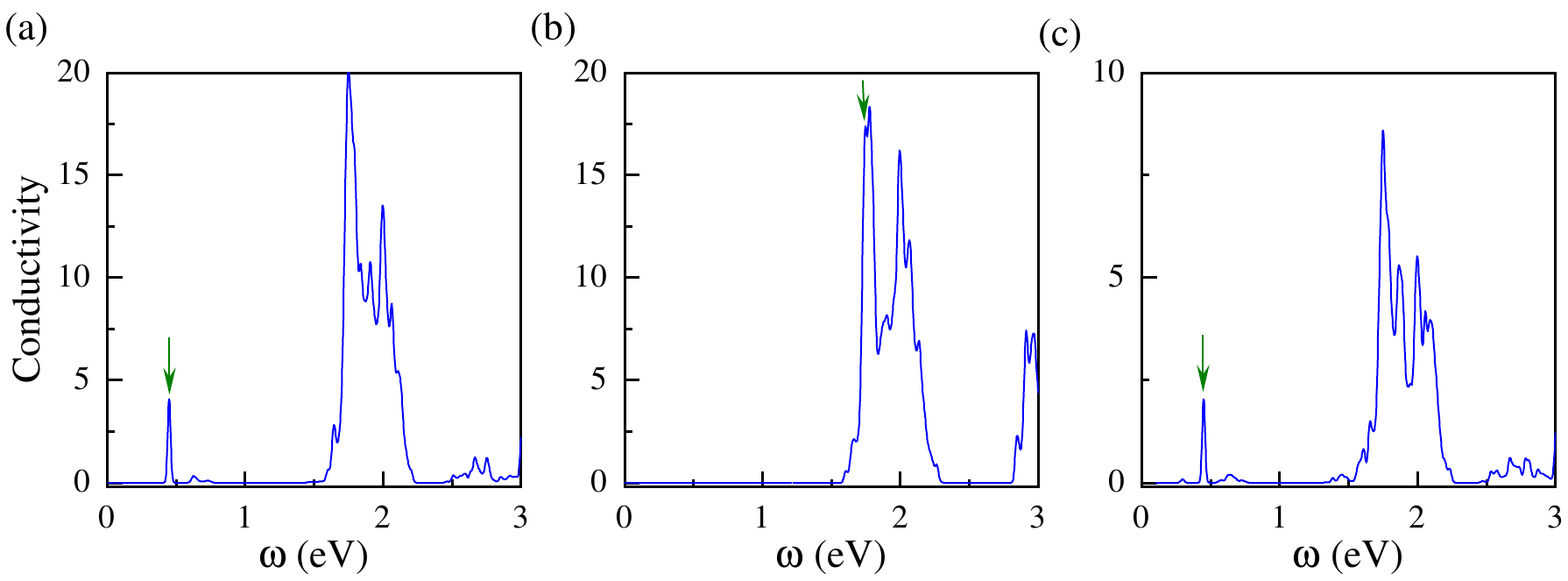}
\caption{Optical conductivity spectra of monolayer CrI$_3$ nanoribbons: (a) Ribbon\_1, (b) Ribbon\_3, and (c) Ribbon\_4. The transition peaks marked by green arrows correspond to the inter-subband transitions shown in Fig. 4.}
\end{figure*}

Additionally, we investigated the optical transition behaviors of CrI$_3$ nanoribbons with different edge configurations. Since Ribbon\_1 and Ribbon\_2 have similar band structures, differing only in the number and degeneracy of edge states, the positions of the transition peaks in their optical conductivity spectra are identical, with only differences in the relative heights of the peaks. Therefore, Fig. 6 only shows the optical conductivity spectra of Ribbon\_1, Ribbon\_3, and Ribbon\_4. The first transition peak marked by the green arrow corresponds to the transition process indicated by arrows in the band structure of Fig. 4.

First, in all calculations of the optical transition spectra for nanoribbons, the Fermi level is set in the middle of the edge states or between the conduction and valence bands, as indicated by the black dashed lines in Fig. 4. For Ribbon\_1 [Fig. 4(a)], the first transition peak originates from transitions between the two edge states, with a band gap of approximately 0.47 eV. Consequently, the energy position of the first transition peak at $\hbar\omega$=0.47 eV corresponds directly to the band gap of the material. This correspondence provides a practical method for determining the band gap and extracting other band parameters from experimental optical spectra. Additionally, due to the large band gap between the edge states and bulk states, no transition peaks appear in the energy range of $\hbar\omega = 0.8-1.5$ eV. Ribbon\_3, which lacks edge states and has a sizable band gap of about 1.6 eV, shows no low-energy transition peaks [[Fig. 6(b)], consistent with its insulating nature. Similarly, Ribbon\_4 has four non-degenerate edge states. When the Fermi level is set in the middle of the edge states, the first transition peak arises from transitions between these edge states.

A comparison of the optical transition spectra for the three distinct nanoribbons in Fig. 6 demonstrates that the energy range of optical transitions can be effectively tuned through edge design, paving the way for designing CrI$_3$-based nanoribbon photoelectric devices.

\section{Conclusion}
In this work, we use a 28-band TB model for monolayer CrI$_3$, combined with linear response theory and the Kubo formula, to systematically investigate the modulation of its electronic band structure and optical conductivity by a vertical electric field. We further explore the influence of different edge configurations on the electronic structure and optical transitions of CrI$_3$ nanoribbons.

Our research results demonstrate that vertical electric fields can selectively modulate the electronic structure of spin channels, exhibiting significant spin-asymmetric responses, which provides a theoretical basis for electric-field-controlled spin-filtering devices. The optical transition behavior is highly dependent on the electric field strength, which allows for dynamic tuning of the peak positions for transitions involving different spin channels. The edge structure of nanoribbons plays a decisive role in their electronic states and optical responses, and these edge states primarily originate from the \emph{d}-orbitals of Cr atoms.

By tailoring edge structures, optical transitions can be effectively engineered across a wide energy range. This study holds great potential for developing advanced nano-optoelectronic devices, including spin filters and magneto-optical modulators, from low-dimensional CrI$_3$. The tight-binding model and linear response method established in this study combine computational efficiency with physical flexibility, overcoming the limitations of first-principles calculations in high-throughput screening and external field modulation, thereby providing an extensible theoretical framework for the application of one-dimensional magnetic materials in spin-optoelectronics.

\section{ACKNOWLEDGMENTS}
This work was supported by the National Natural Science Foundation of China (Grant Nos. 12174100, 12374071 and 12504072), and the Natural Science Foundation of Hunan Province (Grant No. 24C0389).

\end{document}